# Shortwave DPS-QKD Employing a SiN Micro-Ring Resonator as Compact Quantum State Analyser

Florian Honz[(1)], Paul Müllner[(1)], Michael Hentschel[(1)], Stefan Nevlacsil[(1)], Jochen Kraft[(2)], Martin Sagmeister[(2)], Philip Walther[(3)], Rainer Hainberger[(1)], Bernhard Schrenk[(1)]

[(1)] AIT Austrian Institute of Technology, 1210 Vienna, Austria. Author e-mail: florian.honz@ait.ac.at
[(2)] ams-OSRAM AG, 8141 Premstaetten, Austria.  [(3)] Univ. of Vienna, Faculty of Physics, 1090 Austria.

**Abstract** *We show simplified DPS-QKD using a SiN micro-ring resonator operated at 852 nm. A raw-key rate of up to 25.3 kb/s is reached at a QBER suitable for secure-key extraction. Short-reach QKD operation is maintained for zero-touch link layouts with C-band telecom fiber.* ©2024 The Author(s)

## Introduction

Recently, the first quantum computer with more than 1000 physical qubits has been demonstrated [1], in pair with a new error-correction code that proves to be 10-times more efficient than its predecessors [2]. In this imminent age of quantum computation, the need to protect data through an information-theoretic secure approach becomes more pressing. Quantum key distribution (QKD) provides such means by leveraging the quantum properties of light to securely generate a secret key. However, commercial QKD still relies on 19"-sized rack equipment. This calls for a disruptive miniaturization of QKD subsystems to reduce size and cost while further opening new markets for QKD, such as found in short-reach networks. Photonic integrated circuits (PIC) are considered an enabler to do so; However, multi-functional photonic integration platforms are optimized for C-band telecom applications [3]. QKD thus has to strike a balance between chip-scale C-band realizations and shortwave performance gains due to highly efficient and fast room-temperature silicon single-photon detectors [4], which is also paired with an excellent robustness of the QKD signal to co-existing classical data channels [5].

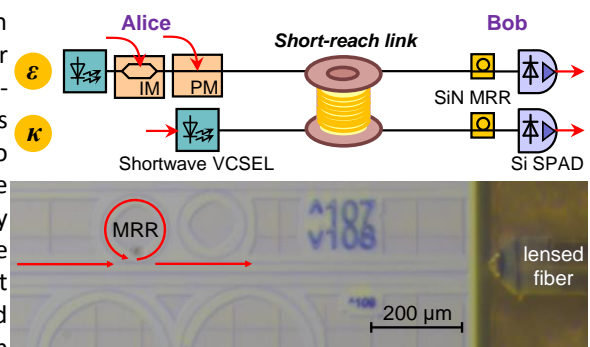

**Fig. 1:** Simplified point-to-point shortwave DPS-QKD link and compact SiN MRR as quantum state analyzer at Bob.

In this work we combine both advantages by demonstrating shortwave (852 nm) differential phase-shift (DPS) QKD including a compact SiN-based micro-ring resonator (MRR) as a quantum state analyser. QKD operation is demonstrated below the quantum bit error ratio (QBER) threshold for secret-key generation at 25.3 kb/s and maintained for few-mode transmission over C-band telecom fiber. We further investigate the simplification of the QKD transmitter by performing quantum state preparation through an energy-efficient directly modulated vertical cavity surface emitting laser (VCSEL) rather than bulky LiNbO$_3$ modulators.

## Simplified Short-Reach DPS-QKD

The peculiarity of QKD systems is often reflected in an increased complexity and cost associated to their constituent elements. Unlike metro-scale installations, where expenditures for QKD deployment can be shared among many network end-users, the cost of cooled C-band single-photon detectors (SPAD) is seen as prohibitive for applications targeting short-reach point-to-point links where no cost sharing can be applied. The tolerable loss for short-reach silica fiber links (~0.2 dB / 100m at 850 nm) and the availability of highly efficient and uncooled Si SPAD technology renders shortwave-QKD as a possible solution; However, integrated photonic shortwave components are greatly missing.

Here, we demonstrate PIC-enabled 850-nm DPS-QKD [6], which encodes individual bits in the phase difference of consecutive photons rather than in the photon state in combination with basis choice at the receiver. We substitute bulky asymmetric Mach-Zehnder interferometers (MZI) employed for quantum state analysis at Bob's receiver with a compact shortwave SiN MRR (Fig. 1), similar as demonstrated for C-band DPS-QKD using a silicon-on-insulator MRR and InGaAs SPAD [7]. In a second step, we will pursue simplification of Alice' transmitter by replacing active external phase modulation [8] or passive path choice in an unbalanced MZI [9] by a low-drive chirp-modulated VCSEL.

## Experimental Shortwave DPS-QKD Setup

Alice will use two different types of transmitters: The first one employed a DFB laser at $\lambda_{Q1}$ = 852 nm which was tuned to a MRR resonance, followed by a Mach-Zehnder Modulator (MZM) for pulse carving and a phase modulator (PM) for state encoding (Fig. 2(a), $\varepsilon$) at $R_{sym}$ = 0.5 Gbaud. The second transmitter was a directly modulated VCSEL at $\lambda_{Q2}$ = 847 nm ($\kappa$). These encoders were driven by an arbitrary waveform generator (AWG). A variable optical attenuator (VOA) then sets the launch power at an average photon number of $\mu_Q$ = 0.1 photons/pulse, while a second VOA emulates the optical budget of the transmission link between Alice and Bob.

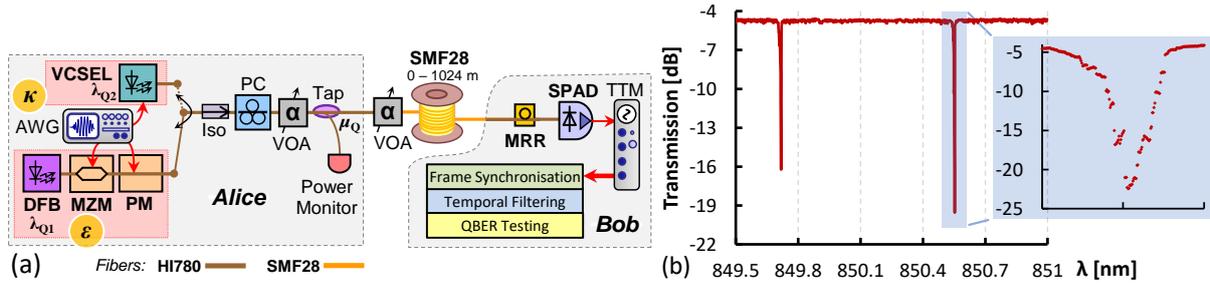

**Fig. 2:** (a) Experimental shortwave DPS-QKD setup and utilized single-mode fibers. (b) Through-port MRR transmission.

Signal transmission has been investigated for a back-to-back (b2b) scenario with a 10-m long HI780 shortwave single-mode fiber and for transmission over ITU-T G.652B compatible C-band single-mode fiber (SMF28) with a length ranging from 256 to 1024 m, resembling a zero-touch short-reach QKD deployment scenario over brownfield links populated with standard telecom fibers. Both fibers feature an increased loss of ~2 dB/km at 850 nm.

At Bob, the optical signal is coupled to the SiN MRR via lensed HI780 fibers before the demodulated signal is detected via a Si SPAD. The MRR was manufactured on a SiN platform and features a radius of 75 µm (Fig. 1). It was designed by means of a semi-analytical approach combining Eigenmode simulations [10] with analytical coupling theory [11]. The gap width between bus and ring was optimized to obtain critical coupling for a loss of 0.5 dB/cm. We achieved a minimum transmission loss of 4.6 dB for fiber-to-fiber coupling over its through-port and the static extinction at its resonance was 18 dB for horizontal polarization (Fig. 2(b)). The free spectral range (FSR) is 346.4 GHz and the FWHM bandwidth of the through-port notch is ~1 GHz. A polarization controller at Alice addresses polarization dependency of the MRR.

The detection events of the SPAD were recorded by a time-tagging module (TTM). We then performed frame synchronization, temporal filtering at 20% of the symbol width, and a real-time raw-key rate and QBER evaluation. Two SPADs were used: *(i)* a green-enhanced Si SPAD-**G** with a dead-time of 77 ns, a dark count rate of 212 cts/s and a detection efficiency of ~9.5% at 850 nm, which is comparable to an InGaAs SPADs operating at 1550 nm (typically featuring 300–600 darks/s, a detection efficiency of 10% and a dead-time of 25 µs), and *(ii)* a red-enhanced Si SPAD-**R** with a dead-time of 50 ns, a dark count rate of 611 cts/s and a detection efficiency of 42% at 850 nm.

**QKD Performance and Few-Mode Operation**

Figure 3(a) presents the raw-key rate and QBER performance as function of the optical budget for the b2b link. At a budget of 0 dB, we reached raw-key rates of 5.8 kb/s and 25.3 kb/s at a QBER of 4.06% and 4.01% for SPAD-**G** and -**R**, respectively. In both cases the QBER is below the QBER threshold of 5% [12], which we consider the limit for secure-key extraction for our coherent-state DPS-QKD implementation. The ~4.4-fold higher detection efficiency of the red-enhanced SPAD-**R** translates directly to an increased raw-key rate. Since the detection efficiency of SPAD-**G** is comparable to C-band InGaAs SPADs, this clearly proves the advantage of the increased detection efficiency of the shortwave SPAD-**R**, together with the generally reduced dead-time of silicon detectors. The key rate can be further improved by optimizing the bandwidth / FSR ratio of the MRR towards the QKD symbol rate $R_{sym}$, to avoid operating this DPS demodulator in the deep notch which causes ~3 dB of additional loss.

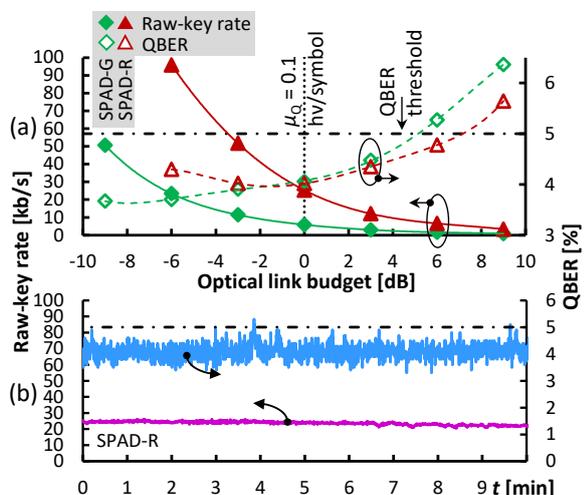

**Fig. 3:** (a) Back-to-Back raw-key rate and QBER over optical link budget, (b) time-resolved QKD performance.

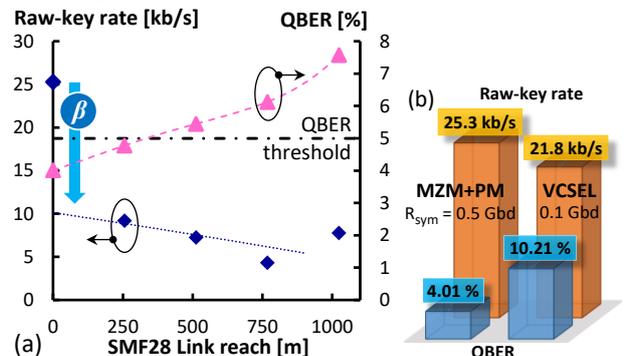

**Fig. 4:** (a) QKD performance after few-mode propagation. (b) Comparison between MZM+PM and VCSEL transmitter.

An optical budget of up to 7.7 dB can be accommodated for the SPAD-**R** configuration before reaching the QBER threshold. For a typical optical interconnect budget of 6 dB, as it is characteristic for intra-datacenter links [13], a raw-key rate of 6.55 kb/s is yielded at a QBER of 4.72%. This allows us to secure a classical channel capacity of 73.3 Gb/s under the NIST limit, which foresees a 256-bit AES key renewal for every 64-GB chunk of data. Figure 3(b) investigates the performance for continuous QKD operation. Both the QBER and raw-key rate remain stable over an acquisition of 10 min, proving a stable spectral MRR setup despite the narrow bandwidth of its through-port resonance.

The transmission performance of the 852 nm quantum channel over standard telecom fiber is investigated in Fig. 4(a). Few-mode propagation applies when operating the shortwave quantum channel over a brownfield link that is populated with a C-band transmission fiber having a cut-off wavelength of 1260 nm. This leads to mode filtering loss due to the inevitable transition from the interconnecting SMF28 fiber (MFD of 10.4 µm @ 1550 nm) to the smaller-core HI780 fiber (MFD of 5 µm @ 850 nm) at Bob's state analyser. This mode filtering effect is evident when comparing to the b2b case: The insertion of a short SMF28 span immediately leads to a sharp drop in raw-key rate (*β*), followed by a slow reduction of the key-rate with increasing SMF28 length. The QBER limit is reached at a SMF28 length of 330 m and a raw-key rate of ~8.2 kb/s. Compared to the b2b measurement, this key-rate corresponds to an optical link budget of 4.8 dB, which translates into a QBER penalty of 0.43% due to few-mode propagation along the C-band transmission fiber.

**Simplified VCSEL-Based QKD Transmitter**

Finally, we investigated possibilities for further complexity reduction of the shortwave QKD link through use of a directly-modulated VCSEL as a greatly simplified and energy-efficient QKD transmitter. The VLI characteristics of the single-mode transistor-outline VCSEL are presented in Fig. 5(a), together with its emission spectrum and its electro-optic modulation response. A direct comparison with the VLI characteristics of the DFB laser (*ε* in Fig. 2(a)) yields the much lower VCSEL bias current of ~3 mA at which a direct AWG drive can be applied, rendering it an energy-efficient solution. The high impedance associated with the single-mode VCSEL design for obtaining a side-mode suppression ratio of 43.9 dB has been addressed through proper RF matching, resulting in an electro-optic bandwidth of 260 MHz. Signal predistortion is then applied as in [14] to accomplish phase encoding through chirp modulation. Figure 5(b) presents the histogram of Alice' frame ($R_{sym}$ = 0.1 Gbaud) as received by Bob. The contrast between 1- and 0-bits is reduced when compared to the ideal LiNbO$_3$ based quantum state encoder, leading to an increased QBER of 10.2% at a raw-key rate of 21.8 kb/s for the b2b case (Fig. 4(b)). The high QBER is attributed to a loss of dynamic extinction during signal demodulation, resulting from the high FM efficiency of the VCSEL, which would require a redesign for the MRR towards a wider bandwidth.

**Conclusion**

We have demonstrated the feasibility of shortwave DPS-QKD for secure key-generation in short-reach links without cost sharing advantage, where a cost-efficient QKD receiver layout is enabled through a compact SiN MRR and a highly efficient Si SPAD. Zero-touch integration in C-band links populated by standard telecom fiber has been validated for a reach up to 330 m. Together with the conceptual simplicity compared to traditional approaches we believe that the shortwave DPS scheme can enable QKD to enter domains deemed as too economically challenging up to this point.

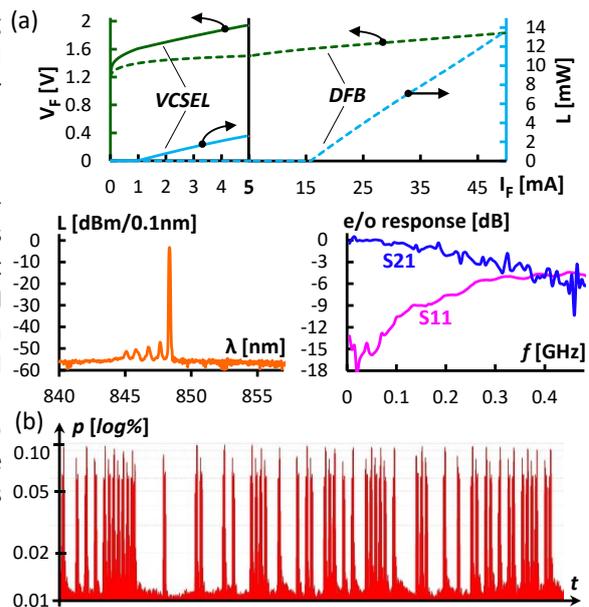

**Fig. 5:** (a) VLI, spectral and e/o characteristics of shortwave single-mode VCSEL. (b) Histogram of Bob's quantum signal.

**Acknowledgements**

This research has received funding from the European Union's Horizon 2020 research and innovation program under grant agreement No. 688173 (OCTCHIP).

## References

[1] D. Castelvecchi, "IBM releases first-ever 1,000-qubit quantum chip," *Nature* [Online], Dec. 2023, https://www.nature.com/articles/d41586-023-03854-1, accessed on 3 April 2024.

[2] S. Bravyi, A. W. Cross, J. M. Gambetta, D. Maslov, P. Rall, and T. J. Yoder, "High-threshold and low-overhead fault-tolerant quantum memory," *Nature*, vol. 627, pp. 778-782, 2024. DOI: 10.1038/s41586-024-07107-7

[3] E. Pelucchi, G. Fagas, I. Aharonovich, D. Englund, E.Figueroa, Q. Gong, H. Hübel, J. Liu, C.-Y. Lu, N. Matsuda, J.-W. Pan, F. Schreck, F. Sciarrino, C. Silberhorn, J. Wang, and K. D. Jöns, "The potential and global outlook of integrated photonics for quantum technologies," *Nature Review Physics*, vol. 4, pp. 194–208, 2022. DOI: 10.1038/s42254-021-00398-z

[4] A. Trenti, M. Achleitner, F. Prawits, B. Schrenk, H. Conradi, M. Kleinert, A. Incoronato, F. Zanetto, F. Zappa, I. Di Luch, O. Çirkinoglu, X. Leijtens, A. Bonardi, C. Bruynsteen, X. Yin, C. Kießler, H. Herrmann, C. Silberhorn, M. Bozzio, P. Walther, H. C. Thiel, G. Weihs, and H. Hübel, "On-Chip Quantum Communication Devices," *Journal of Lightwave Technology*, vol. 40, no. 23, pp. 7485-7497, 2022. DOI: 10.1109/JLT.2022.3201389

[5] M. Ferreira Ramos, M.C. Slater, M. Hentschel, M. Achleitner, H. Hübel, and B. Schrenk, "Datacom-Agnostic Shortwave QKD for Short-Reach Links," in *2024 Optical Fiber Communications Conference and Exhibition (OFC)*, San Diego, CA, USA, 2024, paper Th1C.7.

[6] N. Namekata, H. Takesue, T. Honjo, Y. Tokura, and S. Inoue, "High-rate quantum key distribution over 100 km using ultra-low-noise, 2-GHz sinusoidally gated InGaAs/InP avalanche photodiodes," *Optics Express*, vol. 19, no. 11, pp. 10632-10639, 2011. DOI: 10.1364/OE.19.010632

[7] N. Vokić, D. Milovančev, W. Boxleitner, H. Hübel, and B. Schrenk, "Compact Differential Phase-Shift Quantum Receiver Assisted by a SOI / BiCMOS Micro-Ring Resonator," in *2020 Optical Fiber Communications Conference and Exhibition (OFC)*, San Diego, CA, USA, 2020, paper M4A.4. DOI: 10.1364/ofc.2020.m4a.4

[8] N. K. Pathak., S. Chaughary, Sangeeta, and B. Kanseri, "Phase encoded quantum key distribution up to 380 km in standard telecom grade fiber enabled by baseline error optimization," *Scientific Reports*, vol. 13, no. 15868, 2023. DOI: 10.1038/s41598-023-42445-y

[9] K. Inoue, E. Waks, and Y. Yamamoto, "Differential Phase Shift Quantum Key Distribution," *Physical. Review Letters*, vol. 89, no. 037902, 2002. DOI: 10.1103/PhysRevLett.89.037902

[10] MODE Optical Waveguide & Coupler Solver, Ansys Lumerical Inc., R1.2, 2021.

[11] A. W. Snyder, A. Ankiewicz and A. Altintas, "Fundamental error of recent coupled mode formulations," *Electronics Letters*, vol. 23, pp. 1097-1098, 1987. DOI: 10.1049/el:19870766

[12] E. Waks, H. Takesue, and Y. Yamamoto, "Security of differential-phae-shift quantum key distribution against individual attacks," *Physical Review A*, vol. 73, no. 012344, 2006. DOI: 10.1103/PhysRevA.73.012344

[13] X. Zhou, R. Urata, and H. Liu, "Beyond 1Tb/s Intra-Data Center Interconnect Technology: IM-DD OR Coherent?," *Journal of Lightwave Technology*, vol. 38, no. 2, pp. 475-484, 2020. DOI: 10.1109/JLT.2019.2956779

[14] B. Schrenk, M. Hentschel, and H. Hübel, "Single-Laser Differential Phase Shift Transmitter for Small Form-Factor Quantum Key Distribution Optics," in *2018 Optical Fiber Communications Conference and Exposition (OFC)*, San Diego, CA, USA, 2018, paper Th3E.3. DOI: 10.1364/ofc.2018.th3e.3